\documentclass{PoS}
\title{Simultaneous X-ray/radio observations of Cir X-1}

\ShortTitle{Simultaneous X-ray/radio observations of Cir X-1}

\author{\speaker{Paolo Soleri}\\
        Astronomical Institute ``A. Pannekoek", University of Amsterdam,
	Kruislaan 403, NL-1098 SJ, Amsterdam, The Netherlands\\
        E-mail: \email{psoleri@science.uva.nl}}

\author{{Valeriu Tudose}\\
        Astronomical Institute ``A. Pannekoek", University of Amsterdam,
	Kruislaan 403, NL-1098 SJ, Amsterdam, The Netherlands\\
        E-mail: \email{vtudose@science.uva.nl}}
	
\author{{Rob Fender}\\
        School of Physics and Astronomy, University of Southampton, Highfield,
	Southampton SO17 1BJ\\
        E-mail: \email{rpf@phys.soton.ac.uk}}

\author{{Michiel van der Klis}\\
        Astronomical Institute ``A. Pannekoek", University of Amsterdam,
	Kruislaan 403, NL-1098 SJ, Amsterdam, The Netherlands\\
        E-mail: \email{michiel@science.uva.nl}}

\abstract{We analysed simultaneous X-ray/radio observations of Circinus X-1
collected respectively with RXTE and ATCA in 2000 October and 2002 December
and identified radio flares close to phase 0.0 and 0.5 of the orbital period. To date, there is only
circumstantial evidence for radio flares near phase 0.5. Moreover, in our data set, we clearly associated
both a radio flare and X-ray spectral timing changes with phase 0.0. While for black hole X-ray
binaries the picture of the association between the X-ray and the radio bands is quite well understood,
for neutron star X-ray binaries a clear and complete picture is still missing.}

\FullConference{Bursts, Pulses and Flickering:Wide-field monitoring of the dynamic radio sky\
		 June 12-15 2007\\
		 Kerastari, Tripolis, Greece}

\begin{document}

\section{Introduction}
Low magnetic field neutron star X-ray binaries (NSXBs) are often divided in two classes,
according to their correlated spectral and timing properties: the ``atoll'' and the ``Z''
sources \cite{13}, named after the shape of the track they draw in
the X-ray color-color diagrams (CDs) and the hardness-intensity diagrams (HIDs).

Z sources are brighter than atoll sources and are believed to accrete at
near-Eddington luminosities (0.5-1.0 $L_{Edd}$, \cite{34}). To date seven Galactic
NSXBs have been classified as Z sources: six of them are persistent,
one of them is transient (XTE J1701-462, \cite{16}; as the source faded to quiescence it
switched to atoll-type on the way down, \cite{17}). They are
characterized by three-branched tracks in their CDs and HIDs that in some cases resemble the
character ``Z''. The three branches are called horizontal branch (HB),
normal branch (NB) and flaring branch (FB).
The mass accretion rate, $\dot{m}$, is assumed to drive the transitions between the branches,
increasing monotonically from the HB to the FB, affecting both the spectral and
the timing properties.

Power density spectra (PDS) of the Z sources show several types of quasi periodic
oscillations (QPOs) and noise components \cite{34}, whose presence and
properties are strongly correlated with the position of the source along the Z track
\cite{13}. Three types of low frequency QPOs ($<$100 Hz) are seen in the
Z sources, each of them usually detected in just one of the three branches (but see \cite{35},
\cite{36}); twin kHz QPOs have also been detected in all the Z sources (only marginally detected in
XTE J1701-462, \cite{16}).
Other characteristic noise components of Z sources are the very low frequency noise (VLFN)
and the low frequency noise (LFN).

All the Z-type NS sources are detected in the radio band, showing large and rapid variability, optically
thin and optically thick emission. In 1998 it has been found for the first time (in GX 17+2)
that the radio emission varies as a function of the position in the X-ray CD \cite{27}, decreasing with
increasing mass accretion rate from the HB (strongest radio emission) to the FB (weakest radio
emission). Recently it has been suggested suggested that this behaviour could be universal \cite{21}
(but see \cite{31} for GX 5-1). Extended radio 
jets have been spatially resolved for Sco X-1 and have also
been associated to ultra-relativistic ejections \cite{11}.

\subsection{Cir X-1}
Cir X-1 was discovered in 1971 \cite{20} and has been showing flares with a period of
16.55 days, observed first in the X-ray band \cite{19} and then in the infrared \cite{12}, radio
\cite{14} and optical bands \cite{22}: this fact is interpreted as enhanced accretion
close to the periastron passage of a highly eccentric binary orbit ($e \, \sim \, 0.8$, \cite{23},
\cite{24}). The source is located in the galactic plane at a
distance that has been reported to lie in the range 4-12 kpc (see \cite{18} for
a recent discussion).
Many properties of Cir~X-1 would suggest that this is a black hole candidate (BHC): it has
strong radio emission (e.g. \cite{14}),
ultra-relativistic radio jets (the most relativistic detected so far within our galaxy, \cite{9}),
hard X-ray emission \cite{6} and very strong X-ray variability \cite{26}.
The first strong indication that the binary system harbours a neutron
star has been reported in 1986 \cite{32}, when type-I X-ray bursts have been detected in
EXOSAT data. After \cite{32} no type-I X-ray bursts have been reported.
Shirey et al, in an extensive analysis of RXTE/PCA data, identified typical low-frequency Z-source
features in Cir~X-1 PDS \cite{29} and then a complete  Z track in its high-luminosity
orbital phases \cite{30}. In 1999 it has been noted that the
characteristic timing frequencies of Cir X-1 lie in between those typically associated with NS and BH
systems \cite{28}. Recently twin kHz QPOs have been reported for the first time
in Cir X-1 PDS \cite{3}, a further indication of the nature of the compact object.

Cir X-1 lies within a radio nebula \cite{33} and shows radio jets 
that have been spatially resolved \cite{8}; the radio nebula is 
produced by synchrotron emission likely originating in the interaction 
between the jet and the interstellar medium. Recently 
evidence of an extended arcmin-scale X-ray jet around the source has been reported \cite{15},
in the same direction as the receding radio jet, making Cir X-1 the first secure neutron star
system for which an extended X-ray jet has been resolved.

\section{Observations and data analysis}
\subsection{X-ray data} We analysed 22 RXTE/PCA observations made between 2000
October 1 and 2000 October 26 and 23 RXTE/PCA observations performed between 2002 December 3 and
2002 December 10.

Background subtracted light curves with a time resolution of 16 seconds were obtained from
the ``Standard 2''-mode data, covering the energy range 2-18 keV and dead-time corrections
were applied. We defined two X-ray colors, a  hard color (HC) and a broad color (BC),
in the following energy bands: (8.5-13)/(13-18) keV (HC) and (2-6.3)/(6.3-13) keV (BC).\\
In addition, for each 128-second time interval, we accumulated power density spectra in the
$\sim$2-33 keV and $\sim$2-10 keV energy ranges (respectively for 2000 October and 2002 December),
with a Nyquist frequency of 8192 Hz. Further details on the data analysis will be given in a
future paper (Soleri et al., in preparation).\\
New radio ephemeris (determined in 2007) was used to calculate the orbital phases \cite{25}.

\subsection{Radio data}
We have observed Cir X-1 in radio (simultaneously with X-ray) over multiple epochs on 2000 October 
and 2002 December at 4.8 and 8.6 GHz using the Australia Telescope Compact Array (ATCA). We used 
PKS J1939-6342 (PKS B1934-638) as primary calibrator and PMN J1524-5903 as secondary calibrator
(B1520-58).

\section{Results}
\begin{figure}
\begin{tabular}{c}
\hspace{0.75cm}
\includegraphics[width=12cm]{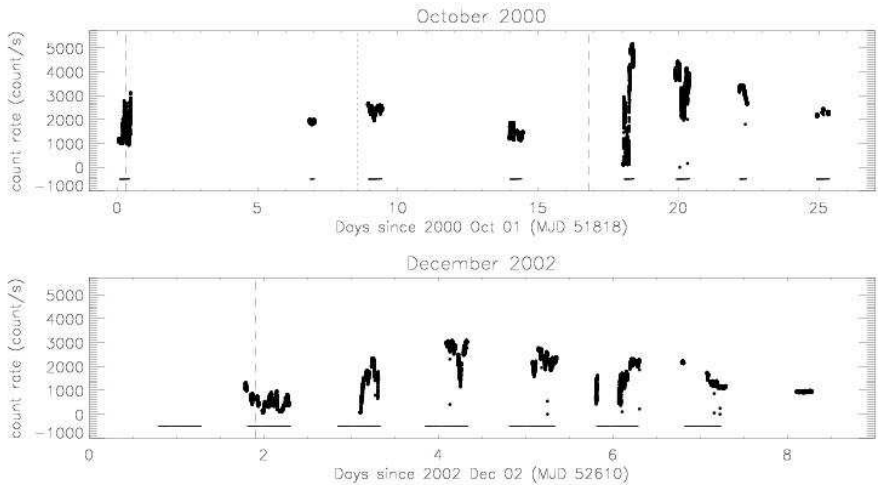}
\end{tabular}
\caption{2-18 keV RXTE/PCA light curves for the two data sets examined here. The bin size is 16
seconds. Only counts from PCU2 were used. Horizontal lines in the bottom part of each light curve
correspond to ATCA radio observations, vertical dashed lines correspond to
phase 0.0 of the orbital period, vertical dotted lines correspond to phase 0.5.}
\label{fig:licu_PCA}
\end{figure}

Figure \ref{fig:licu_PCA} shows the RXTE/PCA light curves of Cir X-1 for our data set: in 2000
October more than an entire orbit was covered, albeit sparsely, while in 2002 December
the data were focused around periastron passage (phase 0.0 of the orbital period).
Since a detailed discussion about the evolution of the X-ray/radio properties in the whole analysed
data set is beyond the scope of this paper, for now we will focus our attention on two particular
orbital phases where we noticed remarkable behaviour. An extensive discussion of the X-ray and
the correlated X-ray/radio properties of this data set will be presented in a future paper
(Soleri et al., in preparation).

\subsection{2002 December - Phase 0.0 of the orbital period}  \label{par:phase0.0}
\begin{figure}
\begin{tabular}{c}
\hspace{1.7cm}
\resizebox{10cm}{!}{\includegraphics[angle=-90]{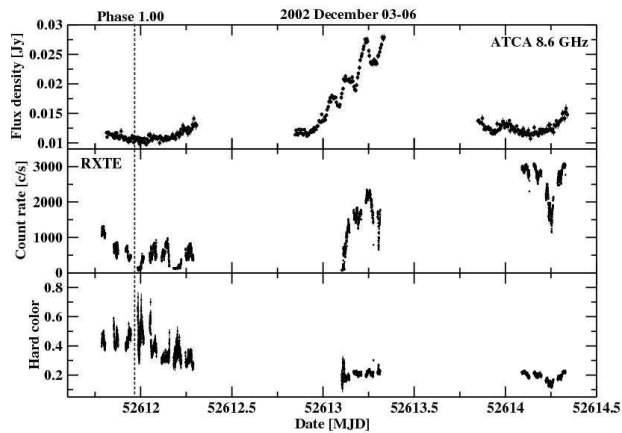}}
\end{tabular}
\caption{Radio and X-ray light curves (top and middle panels) and X-ray hardness curve
(bottom panel) for 2002 December 03-06, between phase 0.99 and 1.15 of the orbital period.
Multiple radio flares are clearly visible after the passage through the periastron}
\label{fig:Dec02_flare}
\end{figure}

\begin{figure}
\begin{tabular}{c}
\includegraphics[width=14cm]{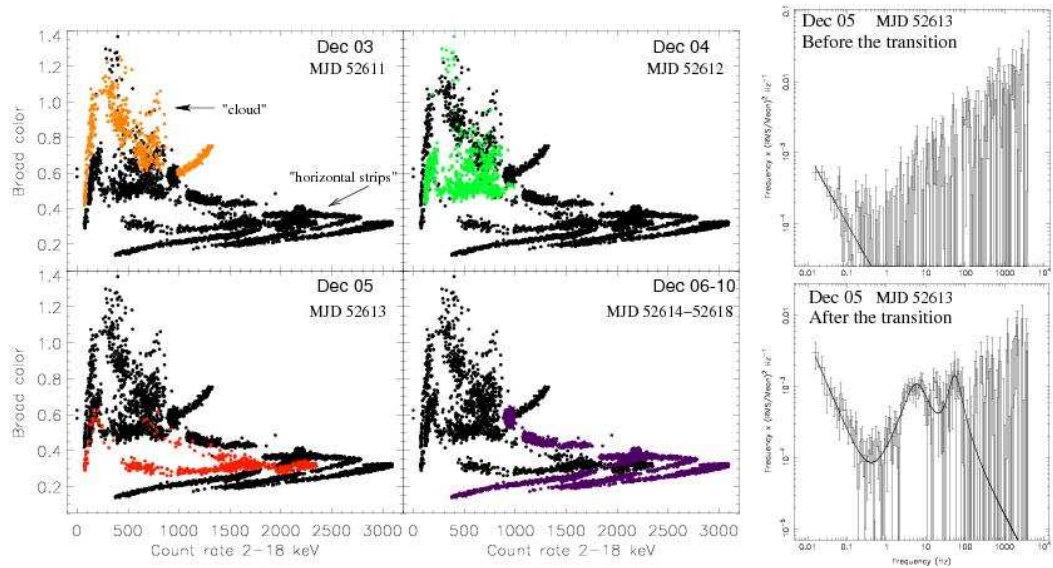}
\end{tabular}
\caption{{\it Left panel:} Hardness-intensity diagram for 2002 December. The bin size is 16 seconds.
In each panel, points corresponding to specific orbital intervals are plotted in different colours.
``Cloud'' and ``horizontal strips" refer to specific zones of the HID (see the text).
{\it Right panel:} power density spectra averaged before (top spectrum) and after (bottom spectrum) the
spectral transition around phase 0.0. Spectra before/after the transition have not been averaged on the whole
cloud/strips region respectively, but on a smaller selection of points (details
will be given in Soleri et al., in preparation). The continuous line represents the best
fit to the data. Spectra are plotted in the $\nu P_{\nu}$ representation \cite{1}.}
\label{fig:hid_Dec2002}
\end{figure}

Figure \ref{fig:Dec02_flare} shows the radio light curve, the X-ray light curve and the X-ray hardness
curve for the orbital phase interval 0.99 - 1.15, in 2002 December. After the passage through
phase 0.0 (but not immediately) a multiple-flare event was clearly detected in radio (both at
4.8 and 8.6 GHz), simultaneous with a sudden increase of the X-ray flux and variations in the X-ray
hardness.  
In Figure \ref{fig:hid_Dec2002} (left panel) we present a X-ray HID where points corresponding to
different days are plotted in different colours: the track drawn by the source in the HID
changes its position and morphology after the passage through phase 0.0 of the orbital phase,
passing from a ``cloud'' located in the left-side of the diagram to a series of horizontal ``strips''
in the bottom side. The transition happens on December 5 (third HID in Figure
\ref{fig:hid_Dec2002} left panel, red points), although it is not sharp: a transition cloud-strips
occurs, followed after approximately 5 hours by a rapid transition (about 200 s) strips-cloud-strips.\\
Power density spectra averaged in the HID before and right after the X-ray spectral transition are shown
in Figure \ref{fig:hid_Dec2002} (right panel). Their properties change considerably: the spectrum
before the transition is characterized by a weak VLFN ({\it rms} = 3.15$\pm$
0.18~\% integrated in the range 0.01-100 Hz) while after the transition the {\it rms} of the VLFN
increases (6.06$\pm$0.58~\% in the same range) and a normal branch oscillation appears ($\nu$ = 5.87$\pm$
0.62 Hz, {\it rms} = 4.48$\pm$0.28~\%).

\subsection{2000 October - Phase 0.5 of the orbital period}  
\begin{figure}
\begin{tabular}{c}
\hspace{1.7cm}
\resizebox{11cm}{!}{\includegraphics[angle=-90]{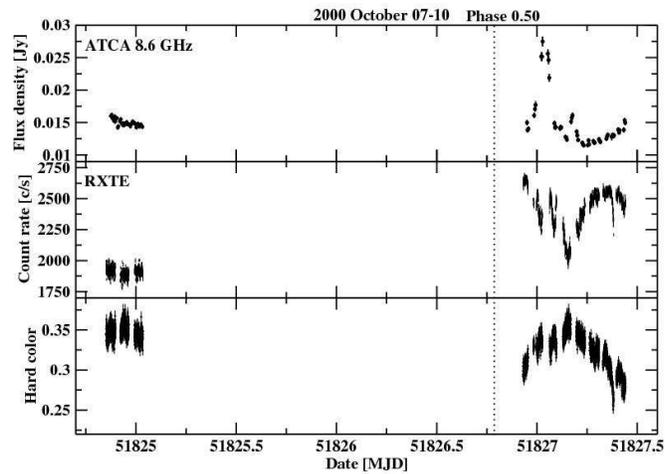}}
\end{tabular}
\caption{Radio and X-ray light curves (top and middle panels) and X-ray hardness curve
(bottom panel) for 2000 October 07-10, between phase 0.40 and 0.55 of the orbital period.
Multiple radio flares are clearly visible close to phase 0.5}
\label{fig:Oct00_flare}
\end{figure}

Figure \ref{fig:Oct00_flare} shows the radio light curve, the X-ray light curve and the X-ray hardness
curve for the orbital phase interval 0.40 - 0.55, in 2000 October.
From the radio light curve, a sequence of radio flares nearby the passage
through the apastron is evident.
Radio flares close to this phase are peculiar and unexpected with respect to what can be found in the
literature where significant radio flux density enhancements are associated
just to phase 0.0 (Tudose et al., in preparation). To date there is only one claim of a
radio flare associated with phase 0.5 \cite{7}.\\
\begin{figure}
\begin{tabular}{c}
\includegraphics[width=14cm]{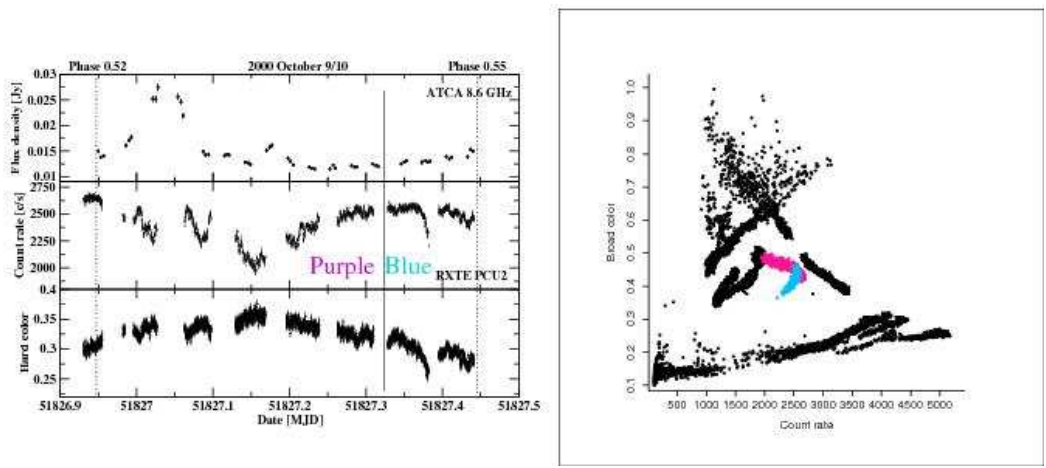}
\end{tabular}
\caption{{\it Left panel:} zoom of Figure 4 around 2000 October 09-10; {\it Right
panel:} HID for 2000 October. Different colours in the HID correspond to different sides
(in respect to the vertical line) of the X-ray light curve}
\label{fig:hid_Oct2002}
\end{figure}
To test whether this sequence is associated with any spectral change in the X-rays, in Figure
\ref{fig:hid_Oct2002} (left panel) we show a ``zoom'' of Figure \ref{fig:Oct00_flare} corresponding to
October 09-10 and the hardness-intensity diagram for all 2000 October (right panel). Purple and blue points
in the HID correspond to the two portions of the light curve in the left panel and are used to mark two
different spectral branches. Since a classification of the spectral branches in the hardness-intensity
diagram is beyond the scope of this paper, for now we will assert without demonstration that the purple
points track a HB and the blue points a NB. A discussion will be presented in a future paper
(Soleri et al., in preparation). On 2000 October 09-10 RXTE clearly caught a spectral transition
between the HB and the NB, however this clearly occurred {\em after} the radio flare. 

\section{Conclusions}
We have analysed simultaneous X-ray/radio observations of Cir X-1 obtained with RXTE and ATCA respectively.

{\bf Phase 0.0:} On 2002 December 05, near periastron passage, we
detected a multiple radio flare event, corresponding to a fast change of the X-ray
spectral/timing properties (in \S \ref{par:phase0.0} we already remarked that in the HID the transition is
not sharp). Our detection could be the first direct observation of a radio flare associated with a spectral 
transition in the X-ray band in a NS system, since no clear association between X-ray spectral/timing
changes and the emission of a radio flare has been reported yet.
Recently it has been suggested that in NS-Z sources the spectral transitions from the HB to the NB 
might be the moment of the launch of transient radio jets \cite{21}, by analogy with what is
suggested for BHCs, where
the hard intermediate state to soft intermediate state transition might drive the ejection of transient
radio plasmons \cite{4}, \cite{10}. In BHCs these state transitions have
been in turn associated with the presence, in the PDS, of a transient low-frequency QPO, the so called type-B
\cite{2}. An association between the type-B QPO detected in BHCs and the normal branch oscillation
observed in Z sources has been suggested in \cite{5}, since these QPOs present
similar properties: on 2002 December we could associate the radio flare to a X-ray spectral
transition that is in turn associated with the presence of a normal branch oscillation, although we do
not see any obvious HB to NB passage.   

{\bf Phase 0.5:} On 2000 October 09 we detected a radio flare near the passage through
the apastron. In the literature there is only circumstantial evidence for  radio flares at phase 0.5 \cite{7},
maybe partly because the observing strategies concentrated around phase 0.0. This is the first clear
evidence of a radio flare near phase 0.5 (Tudose et al., in preparation). In the RXTE data we also
detected a spectral transition on 2000 October 09 about 2.5 hours after the end of the
second radio flare, with the source moving from the HB to the NB. As it has been suggested in \cite{21},
this transition might be the moment of the launch of transient radio jets, but
in our specific case the time difference between the two events does not allow us to associate them.
Unluckily we do not have RXTE data right before the beginning of the radio flaring activity (see Figure
\ref{fig:Oct00_flare}).


\begin{thebibliography}{99}
\bibitem{1} Belloni, T., Psaltis, D., van der Klis, M., 2002, ApJ, 572, 392
\bibitem{2} Belloni, T., Homan, J., Casella, P., van der Klis, M., Nespoli, E., Lewin, W.~H.~G.,
Miller, J.~M., M{\'e}ndez, M., 2005, A\&A, 440, 207
\bibitem{3} Boutloukos, S., van der Klis, M., Altamirano, D., Klein-Wolt, M, Wijnands, R., Jonker,
P.~G., Fender, R.~P., 2006, ApJ, 653, 1435
\bibitem{4} Casella, P., Belloni, T., Homan, J., Stella, L., 2004, A\&A, 426, 587
\bibitem{5} Casella, P., Belloni, T., Stella, L., 2005, ApJ, 629, 403
\bibitem{6} Ding, G.~Q., Qu, J.~L., Li, T.~P., 2003, Astr. J., 596, L219
\bibitem{7} Fender, R.~P., 1997, Proceedings of the Fourth Compton Symposium, ed. Dermer, C.~D., 
Strickman, M.~S, Kurfess, J.~D., p. 798
\bibitem{8} Fender, R., Spencer, R., Tzioumis, T., Wu, K., van der Klis, M., van Paradijs, J., Johnston, H.,
1998, ApJL, 506, L121
\bibitem{9} Fender, R., Wu, K., Johnston, H., Tzioumis, T., Jonker, P., Spencer, R., van der Klis M.,
2004, Nat, 427, 222
\bibitem{10} Fender, R.~P., Belloni, T.~M., Gallo, E., 2004, MNRAS, 355, 1105
\bibitem{11} Fomalont, E.~B., Geldzahler, B.~J., Bradshaw, C.~F., 2001, ApJ, 559, 283
\bibitem{12} Glass, I.~S., 1978, MNRAS, 183, 335
\bibitem{13} Hasinger, G. \& van der Klis, M., 1989, A\&A, 225, 79
\bibitem{14} Haynes, R.~F., Jauncey, D.~L., Murdin, P.~G., Goss, W.~M., Longmore, A.~J., Simons, L.~W.~J., 
Milne, D.~K., Skellern, D.~J., 1978, MNRAS, 185, 661
\bibitem{15} Heinz, S., Schulz, N.~S., Brandt, W.~N., Galloway, D.~K., 2007, ApJL, 663, L93
\bibitem{16} Homan, J., van der Klis, M., Wijnands, R., Belloni, T., Fender, R., Klein-Wolt, M., 
Casella, P., M{\'e}ndez, M., Gallo, E., Lewin, W.~H.~G, Gehrels, N., 2007a, ApJ, 656, 420
\bibitem{17} Homan, J., Wijnands, R., Altamirano, D., Belloni, T., 2007b, ATEL, 1165, 1
\bibitem{18} Jonker, P.~G. \& Nelemans, G., 2004, MNRAS, 354, 355
\bibitem{19} Kaluzienski, L.~J., Holt, S.~S., Boldt, E.~A., Serlemitsos, P.~J., 1976, ApJL, 208, L71
\bibitem{20} Margon, B., Lampton, M., Bowyer, S., Cruddace, R., 1971, ApJL, 169, L23
\bibitem{21} Migliari, S. \& Fender, R.~P, 2006, MNRAS, 366, 79
\bibitem{22} Moneti, A., 1992, A\&A, 260, L7
\bibitem{23} Murdin, P, Jauncey, D.~L., Lerche, I., Nicolson, G.~D., Kaluzienski, L.~J., Holt, S.~S., Haynes,
R.~F., 1980, A\&A, 87, 292
\bibitem{24} Nicolson, G.~D., Glass, I.~S., Feast, M.~W., 1980, MNRAS, 191, 293
\bibitem{25} Nicolson, G.~D., 2007, ATEL, 985, 1
\bibitem{26} Oosterbroek, T., van der Klis, M., Kuulkers, E., van Paradijs, J., Lewin, W., 1995, A\&A,
297, 141
\bibitem{27} Penninx, W., Lewin, W.~H.~G., Zijlstra, A.~A., Mitsuda, K., van Paradijs J., 1988,
Nat, 336, 146
\bibitem{28} Psaltis, D., Belloni, T., van der Klis, M., 1999, ApJ, 520, 262
\bibitem{29} Shirey, R.~E., Bradt, H.~V., Levine, A.~M., Morgan, E.~H., 1998, ApJ, 506, 374
\bibitem{30} Shirey, R.~E., Bradt, H.~V., Levine, A.~M., 1999, ApJ, 517, 472
\bibitem{31} Tan, J., Lewin, W.~H.~G., Hjellming, R.~M., Penninx, W., van Paradijs, J., van der Klis, M.,
Mitsuda, K., 1992, ApJ, 385, 314
\bibitem{32} Tennant, A.~F., Fabian, A.~C., Shafer, R.~A., 1986, MNRAS, 221, 27P
\bibitem{33} Tudose, V., Fender, R., Kaiser, C., Tzioumis, A., van der Klis, M., Spencer R., 2006,
MNRAS, 372, 417
\bibitem{34} van der Klis, M., 2006, in ``Compact Stellar X-ray Sources'', ed. W. H. G. Lewin \&
M. van der Klis, Cambridge University Press, Cambridge, p. 39
\bibitem{35} Wijnands, R.~A.~D., van der Klis, M., Psaltis, D., Lamb, F.~K., Kuulkers, E., Dieters,
S., van Paradijs, J., Lewin, W.~H.~G., 1996, ApJL, 469, L5
\bibitem{36} Wijnands, R.~A.~D., van der Klis, M., Kuulkers, E., Asai, K., Hasinger, G., 1997,
A\&A, 323, 399
\end{thebibliography}
\end{document}